\documentclass[preprint]{aastex}
\begin{document}
\title{
XMM-Newton observations of the dwarf nova
RU Peg in quiescence: Probe of the boundary layer 
}
\author{\c{S}\"{o}len Balman}
\affil{Department of Physics, 
Middle East Technical University, 
Ankara, Turkey}
\email{solen@astroa.physics.metu.edu.tr}  

\author{Patrick Godon\altaffilmark{1}, Edward M. Sion} 
\affil{Department of Astronomy \& Astrophysics, 
Villanova University,
Villanova, PA 19085}
\email{patrick.godon@villanova.edu; edward.sion@villanova.edu}

\author{Jan-Uwe Ness} 
\affil{XMM-Newton Science Operations Centre, 
European Space Agency (ESA/ESAC), 
28691 Villanueva de la Canada, Madrid, Spain} 
\email{juness@sciops.esa.int} 

\author{Eric Schlegel} 
\affil{Department of Physics \& Astronomy,  
University of Texas at San Antonio, 
San Antonio, TX 78249} 
\email{eric.schlegel@utsa.edu} 

\author{Paul E. Barrett} 
\affil{United States Naval Observatory, 
Washington, DC 20392} 
\email{barrett.paul@usno.navy.mil} 

\author{Paula Szkody} 
\affil{Astronomy Department,   
University of Seattle, 
Seattle, WA 98195}  
\email{szkody@astro.washington.edu} 
 
\author{ } 
\affil{ } 

\altaffiltext{1}
{
Visiting at the Johns Hopkins University, Baltimore, MD 21218
}

\begin{abstract}

We present an analysis of X-ray and UV data  
obtained with the XMM-Newton Observatory 
of the  long period dwarf nova RU Peg. 
RU Peg contains a massive white dwarf, 
possibly the hottest white dwarf in a dwarf nova, it has a  
low inclination, thus optimally exposing its  
X-ray emitting boundary layer,  and has an excellent 
trigonometric parallax distance. 

We modeled  the X-ray data using  XSPEC  
assuming a multi-temperature
plasma emission model built from the MEKAL code (i.e., CEVMKL).
We obtained a maximum temperature of 31.7 keV, based on the
EPIC MOS1, 2 and pn data, indicating that 
RU Peg has an X-ray spectrum harder than most dwarf novae, except U Gem. 
This result is consistent with and indirectly confirms the large 
mass of the white dwarf in RU Peg. 
The X-ray luminosity we computed corresponds to a boundary layer luminosity
for a mass accretion rate of $2 \times 10^{-11}M_{\odot}$/yr (assuming
$M_{wd}=1.3M_{\odot}$),
in agreement with the expected quiescent accretion rate. 
The modeling of the O\,{\sc viii} emission line at 19\AA\  as observed by the 
RGS implies a projected 
stellar rotational velocity $v_{rot} \sin{i}=$695 km$~$s$^{-1}$, i.e. the line is emitted from
material rotating at $\sim$936-1245 km$~$s$^{-1}$ ($i \sim 34^{\circ}-48^{\circ}$) or about 1/6
of the Keplerian speed; 
this velocity is much larger than the rotation speed of the white dwarf
inferred from the Far Ultraviolet Spectroscopic Explorer ({\it FUSE}) spectrum. 
Cross-correletion analysis yielded an undelayed (time lag $\sim$ 0) component 
and a delayed component of 116$\pm$17 sec where the X-ray variations/fluctuations
lagged the UV variations. This indicates that the UV fluctuations in the inner disk
are propagated into the X-ray emitting region in about 116 sec. The undelayed component may be
related to irradiation effects.

\end{abstract}

\keywords{accretion, accretion disks, binaries: close, 
Stars: white Dwarfs, Stars: dwarf novae (RU Peg), X-rays: binaries}  

\section{Introduction}
\subsection{The long period dwarf nova RU Peg} 

Dwarf novae (DNe) are a class of weakly-magnetic cataclysmic variables 
(CVs) which are interacting compact binaries  
in which a white dwarf (WD, the primary star) accretes matter 
and angular momentum from a main (or post-main) sequence star (the secondary) 
filling its Roche-lobe. The matter is transferred, at a continuous or 
sporadic {\it accretion} rate ($\dot{M}$), by means of an {\it accretion} disk
usually reaching all the way to the WD surface. Ongoing accretion at 
a low rate (quiescence) is interrupted every few weeks to months by 
intense accretion (outburst) of days to weeks (a dwarf nova accretion 
event).
DNe are powerful X-ray sources with luminosities of 
$10^{30}-10^{33}$erg$~$s$^{-1}$. The X-ray emission is believed to
originate in the boundary layer (BL) between the slowly rotating accreting 
WD and the fast rotating (Keplerian) inner edge of the accretion disk, 
where the material dissipates its remaining rotational kinetic 
energy before it accretes onto the surface of the WD. 
The typical DNe, i.e. those systems exhibiting
normal DN outbursts are the U Gem sub-type of DNe 
(according to the classification of  
\citet{rit03}) and  are located  above the period gap
($P_{orb}>3$hr).  

RU Peg is a U Gem type DN with an orbital period $P_{orb}$ = 8.99 hrs, a secondary spectral type K2-5V, a 
primary (WD) mass $M_{wd} = 1.29_{-0.20}^{+0.16} M_{\odot}$, 
and a secondary mass $M_{2} =0.94\pm0.04 M_{\odot}$ 
\citep{sto81,wad82,sha83}. 
The system has a magnitude range $V_{max} - V_{min} \approx 9.0 - 13.1$   
with outbursts lasting $\sim$20 days and recurring every $\sim$50 days. 
The near-Chandrasekhar WD mass has been corroborated by the sodium (8190\AA)
doublet radial velocity study of \citet{fri90}.  They obtained a mass of
1.38$\pm$0.06 M$_{\odot}$ for the    WD      and also found a range of inclination
angles between 34$^{\circ}$-48$^{\circ}$
in agreement with the range of plausible
inclinations found in the study by \citet{sto81}. 
More recently, a Hubble Fine Guidance Sensor ({\it FGS}) 
parallax of $3.55\pm0.26$ mas was measured by \citet{joh03} yielding
a distance of $282 \pm 20$ pc.

RU Peg was observed with IUE under several different observing 
programs both in quiescence and in outburst 
and was part of several survey-like studies  
(e.g. \citet{lad85,ver87,szk91} to cite just a few).  

\citet{sio02} modeled  4 IUE spectra obtained in deep
quiescence with accretion disks and photospheres. 
They found that a very  hot WD dominated the FUV spectrum 
with a temperature T$_{eff} = 50-53,000$K which places 
RU Peg among the hottest WDs in DNe.              
The distance corresponding to their best fitting, high gravity
($Log(g)=8.7$) photosphere models was 250 pc.

More recently, \citep{god08} modeled the {\it{FUSE}} spectrum of RU Peg
in quiescence and obtained a WD with a temperature of 70,000K,  
a rotational velocity of 40 km$~$s$^{-1}$, assuming $Log(g)=8.8$ and  
a distance of 282 pc. In this later study the higher temperature obtained  
in the model fitting is mainly a consequence of the assumed larger distance 
and gravity.  It is clear that 
RU Peg has a  massive WD and therefore a deep potential well, 
and its surface temperature is very large ($>$50,000K), 
possibly pointing to strong accretional and boundary
layer heating.  
For these reasons, we chose RU Peg as our X-ray target, 
as it is expected to be a copious source of X-rays  
and should be an ideal candidate to study its boundary layer.

\subsection{The Boundary Layer}

The boundary layer is that region between the slowly rotating accreting 
WD and the fast rotating (Keplerian) inner edge of the accretion disk.  
For accretion to occur, gravity has to overcome the centrifugal acceleration, 
and this happens in the boundary layer    
as the material dissipates its remaining rotational kinetic 
energy before being accreted onto the surface of the WD. In the
boundary layer the rotational velocity is sub-Keplerian and decreases
inwards.   
 
The standard disk theory \citep{sha73} 
predicts a BL luminosity  almost 
equal to the disk luminosity \citep{pri81} : 
\begin{equation} 
L_{BL} = ( 1 - \beta^2) L_{disk} = ( 1 - \beta^2)  
\frac { GM_* \dot{M}}{ 2R_*}, 
\end{equation} 
where $G$ is the gravitational constant, $M_*$ is the mass
of the accreting star, $R_*$ its radius, $\dot{M}$ is the mass
accretion rate, and $\beta$ is the stellar angular velocity
in Keplerian units $\beta  =\Omega_* /\Omega_K(R_*)$. For most
systems the WD rotational velocity is of the order of a few $100$ km$~$s$^{-1}$ and
one therefore has $\beta << 1$ and $L_{BL} \approx L_{disk}$. 
For a star rotating at (e.g.) 10\% of the breakup velocity eq.(1) gives  
$L_{BL}= 0.99 L_{disk}$.  However, the standard disk theory 
does not take into account the boundary layer explicitly.  
Using a one-dimensional approach, \citet{klu87} has shown that  
part of the BL (kinetic) energy actually goes into the spinning of the star
in the equatorial region, and the BL luminosity is more accurately
given by the relation  
\begin{equation} 
L_{BL} = \left( 1 - \beta \right)^2 L_{disk} . 
\end{equation}
For a star rotating at 10\% of the breakup velocity, eq.(2) gives 
$L_{BL} = 0.81 L_{disk}$, which is significantly different than eq.(1).

Because of its small radial extent, the BL is expected to emit
its energy in the X-ray bands ($L_{BL} \approx L_{X-ray}$). 
At high accretion rate (or during dwarf nova outburst), the BL is
expected to be optically thick with a temperature 
$T \approx 10^5-10^6$K \citep{god95,pop95}, 
and emits in the soft X-ray band. 
Observations of CVs in high state 
(e.g. \citet{mau95,bas05})
confirm these predictions. Accretion onto white dwarfs in some symbiotics 
presents similar soft X-ray emissions \citep{lun07,lun08,ken09} 
because of the large mass accretion rates in these systems.     
At low mass accretion rate (or during
dwarf nova quiescence), as the density is much decreased, the BL 
becomes optically thin and emits in the hard X-ray band with 
$T\approx 10^8$K \citep{nar93} (see below for observational evidence
of such optically thin BLs). 
Hence, during quiescence the emission should 
arise from a very hot plasma very close to the WD surface. 

Previous X-ray observations of dwarf nova systems in quiescence
(e.g. \citet{van87,van96,bel91}) while confirming the presence of
hard X-ray, deduced that, contrary to the theory, 
the quiescent BL luminosity was underluminous. 
Namely, they found $L_{BL}= L_{X-ray} << L_{disk}$, which confirmed the
original claim that BLs were actually missing (Ferland et al. 1982). 
However, these earlier results assumed the disk to be the source 
of the optical and ultraviolet radiation ($L_{disk}=L_{opt}+L_{UV}$)
and used eq.(1) rather than eq.(2).
More recently, X-ray Multi Mirror-Newton (XMM-Newton) 
observations of 8 DNe in quiescence 
by \citet{pan03,pan05} revealed  that a significant part of the emitted 
FUV flux ($L_{UV}$) actually originates from the WD itself, 
and the evidence for underluminous BLs in 
quiescent DNe was refuted. 
For RU Peg, {\it FUSE} observations \citep{god08}  
indicate that the WD contributes possibly most of the
FUV light with a temperature $T>50,000$K. 
In addition, \citet{god05} noted that the region where the boundary
layer meets the inner edge of the Keplerian disk \citep{pop99} can also
contribute some FUV flux. 
It is clear now that one cannot just compare 
the X-ray luminosity to the optical + UV luminosity to check
whether the boundary luminosity is as large as the disk luminosity. 

For the 8 dwarf novae caught in quiescence, 
\citet{pan03,pan05},  
using the XMM-Newton data,  
obtained X-ray boundary layer luminosities 
of the order of $\sim 1 \times 10^{31}$ergs$~$s$^{-1}$ to $6.6 \times 10^{32}$ergs$~$s$^{-1}$, 
with temperatures ranging from $\sim 8$ to $55$ keV, 
and mass accretion rates deduced from X-rays in the range 
$10^{-12}M_{\odot}$/yr to $10^{-10}M_{\odot}$/yr.   
In this work, we present XMM-Newton observations  
of RU Peg taken in quiescence to derive its X-ray luminosity
and gain information on its boundary layer. 

\section{Observations and Analysis} 

The XMM-Newton Observatory \citep{jan01} has three 1500 cm$^2$
X-ray telescopes each with an European Photon Imaging Camera (EPIC)
at the focus. Two of the telescopes have Multi-Object Spectrometer (MOS) CCDs
\citep{tur01} and the last one uses pn CCDs  \citep{str01}
for data recording. There are two Reflection Grating Spectrometers
(RGS) \citep{den01}. The Optical Monitor (OM), a photon counting instrument, 
is a co-aligned 30-cm optical/UV telescope, providing for the first time the 
possibility to observe simultaneously in the X-ray and optical/UV regime from a single platform \citep{mas01}.

RU Peg was observed (pointed observation) with XMM-Newton 
for an duration of 53.1 ks on June 9th, 2008, at 07:16:50.0
UTC (obsID 0551920101). At that time the system was at a visual magnitude of
$\sim  12.5$, about 2 months into quiescence and 2 weeks before 
the next outburst (from AAVSO data \footnote{http://www.aavso.org/}). 
Data were collected with the 
EPIC MOS and pn cameras in the prime partial window2 and prime full window imaging mode,
respectively,
the Reflection Grating Spectrometer 
and the Optical Monitor using the fast imaging mode ($\le$0.5 sec time
resolution)  with the UVW1 filter (240-340 nm).

We analysed the pipeline-processed data using Science Analysis Software (SAS)
version 9.0.0. Data
(single- and double-pixel events, i.e., patterns 0--4 with Flag=0 option for pn and patterns $\le$ 
12 with Flag=0 for MOS1,2) were extracted from
a circular region of radius 60$^{\prime\prime}$ for pn, and 45$^{\prime\prime}$ MOS1 \& MOS2
in order to perform spectral analysis together with the background events 
extracted from a
source free zone normalized to the source extraction area.
We checked the pipeline-processed event file for any existing flaring
episodes and no sporadic events in the background were detected with count rate higher than
0.08 c s$^{-1}$ (for MOS1,2) and 0.5 for pn detectors. 
Table 1 displays the background subtracted count rates
for the EPIC pn, MOS1 and MOS2. 

The RGS observations were carried out using the standard spectroscopy mode for 
readout.  
We reprocessed the data using the XMM-SAS routine {\sc rgsproc}.
We first made event files and determined times of
low background from the count rate on CCD 9 (which is closest to the optical axis).
The final exposure times and net count rates
showed that there were no sporadic high background events in our data.
Table 1 displays the background subtracted count rates for RGS1 and RGS2.
Source and background counts for the RGS were extracted using the standard spatial 
and energy filters for the source position,
which defines the spatial extraction regions as well as the wavelength zero point.
\\

\subsection{X-ray and UV Light curves} 

The UV and EPIC pn X-ray light curves are shown together for comparison
in Figure 1, where we have scaled the UV count rate to fit the X-ray count rate
for a better comparison. The count rate for the UV data ranges between $\sim$50 c/s 
and $\sim$100 c/s with a time average of $\sim$71.8 c/s. 
For UV-bright objects, a count rate of 1 s$^{-1}$ in the UVW1 filter translates into 
a flux of 
$4.5 \times 10^{-16}$ergs$~$cm$^{-2}$s$^{-1}$\AA$^{-1}$ at 290 nm (see the online
XMM-Newton 
documentation\footnote{
http://xmm.vilspa.esa.es/external/xmm\_user\_support/documentation/uhb/index.html}).  
This gives for RU Peg a flux of 
$\approx 2.2-4.5 \times 10^{-14}$ergs$~$cm$^{-2}$s$^{-1}$\AA$^{-1}$,
corresponding to a luminosity of  
$\approx 2.1-4.3 \times 10^{29}$ergs$~$s$^{-1}$\AA\ $^{-1}$ (at a distance d=282pc). 
The time modulation of the UV data follows closely the 
time modulation of the X-ray data except around 
$t\approx $12 ks, 20-22 ks, 31 ks, \& 48.5 ks, where the
UV has {\it relatively} more flux for a duration of several hundred seconds
(and up to 1,000 s). 
Since the UV is expected to be emitted further out than the X-rays,
these four epochs where the UV light curves do not decrease
as much as the X-rays might be due to the occultation of
the X-ray emitting material by the WD, while the UV
emitting region is not hidden from the observer. 

In order to study the correlation between the X-ray and the UV variability,
we calculated the cross-correlation between the two light curves. 
We used time bins of 5 sec averaging several power spectra with 128
bins for the analysis. The  resulting correlation coefficient 
as a function of time lag is shown in Figures 2a \& 2b. 
The correlation coefficient is normalized to have a maximum value of 1.
The curve shows a clear asymmetry indicating existence of time delays.
We, also, detect a strong peak
near zero time lag suggesting a significant correlation between X-rays
and the UV light curves. We call this the undelayed component (see Figures 2a
\& 2b).
The positive time-lag in the asymmetric profile shows that the 
X-ray variations are delayed relative to those in the UV.
In order to calculate an average time-lag that would produce the asymmetric
profile, we fitted the varying cross-correlation by two lorentzians,
with time parameter fixed at 0.0 lag and the other set as free. The resulting fit 
yields a lag of 116$\pm$17 sec. This is the delayed component.

\subsection{EPIC Spectrum} 

We performed spectral analysis of the EPIC data using the SAS task
{\sc ESPECGET}
and derived the spectra of the source and the background together with the appropriate response
matrices and ancillary 
files. How the photons were extracted is described in Section 2.
The EPIC pn, EPIC MOS1 and EPIC MOS2 spectra were simultaneously fitted
to derive the spectral parameters. The spectral analysis was performed
using XSPEC version 12.6.0q (Arnaud 1996). A constant factor
was included in the spectral fitting to allow for a normalization
uncertainty between the EPIC pn and EPIC MOS instruments.
We grouped the pn and MOS spectral energy channels so that there is a minimum of 
80 (MOS1,2)-150 (pn) 
counts in a bin to improve the statistical quality of the spectra.
The fits were conducted in the 0.2-10.0 keV range.
The simultaneously fitted spectra from the three EPIC instruments
are shown in Figure 3. 
We modeled the X-ray spectrum of RU Peg in a similar fashion as \citet{pan05} 
and fitted the data with
(TBabs$\times$CEVMKL) model within XSPEC. TBabs is the Tuebingen-Boulder ISM absorption model
(Wilms, Allen and McCray 2000) and  CEVMKL is a multi-temperature plasma emission model built 
from the mekal code \citep{mew85}. Emission measures follow a power-law in temperature (i.e. emission 
measure from temperature T is proportional to $(T/T_{max})^{\alpha-1}$). The residuals
in Figure 3 show systematic fluctuation around the 6.7-6.9 keV iron line complex
mainly from the EPIC pn data and some small low energy fluctuations exist in the
MOS2 data, as well. This is due to the 
CTI (Charge transfer inefficiency) problem, in the pn (and possibly MOS2) instrument.
These generally occur around lines due to small calibration errors and mostly effect
only the line shapes leaving systematic residuals and increasing the
reduced $\chi^2$ of the fits. Our EPIC MOS1 data does not exhibit any CTI effects
and the reduced $\chi^2$ for the fit to this data alone is 1.15 (d.o.f. 290). 
The reduced $\chi^2$ for the simultaneously fitted  spectra is higher than the
value for the MOS1 fit, but the spectral parameters for all three instruments
are almost the same within the errors. 
Table 2 contains the spectral parameters from fits (using three detectors simultaneously) 
with the (TBabs$\times$CEVMKL) model. 
Errors are given at 90\% conf. level. We find a maximum plasma temperature in a range 
29-33 keV and mostly solar abundances of elements aside from oxygen and neon which we
calculate to be subsolar. The unabsorbed X-ray flux is 4.1$^{+0.2}_{-0.2}$$\times 10^{-11}$ erg$~$s$^{-1}$cm$^{-2}$ 
which translates to a luminosity of 4.1$^{+0.3}_{-0.3}$$\times 10^{32}$ erg$~$s$^{-1}$ at 282 pc (see sec. 1).
The neutral hydrogen column density is 4.3-4.5$\times 10^{20}$ cm$^{-2}$.

\subsection{RGS Spectrum} 

The RGS analysis tool {\sc rgsproc} was used to obtain
RGS1 and RGS2 spectra and to produce fluxed spectrum (i.e., using {\sc rgsfluxer})
The resultant fluxed spectrum 
of the combined RGS1 and RGS2 detectors is shown in Figure 4
with line identifications. The detected lines and corresponding
wavelengths are listed in Table 3. Fluxed spectra are obtained just by dividing the count spectrum by the RGS
effective area. It  neglects the redistribution of monochromatic response into the dispersion channels.
Since  proper response is not utilized by the fluxed spectrum, 
in order to perform a similar spectral
fit/analysis using the same (TBabs$\times$CEVMKL) model within XSPEC,
we used the count rate spectra produced for RGS1 and RGS2 simultaneously
with the appropriate response files for each detector. This is
a very efficient approach to find the spectral parameters 
in comparison with the EPIC results.  
The fitted RGS1 and RGS2 spectra are shown in Figure 5.
Table 4 contains the spectral
parameters from the fit with the (TBabs$\times$CEVMKL) model within XSPEC. 
Errors are 90\% conf. level and the fit is performed between 0.2-2.5 keV.

The maximum temperature from the fits with the RGS data yield 24 keV  
with a large error range of  17-41 keV since the spectrum has a lower count rate 
and a lower upper energy boundary (i.e., 2.5 keV)  
compared with the EPIC data (i.e., 12 keV). We find that most of the RGS spectral 
parameters are  consistent for the EPIC results. Oxygen and neon are subsolar 
in abundance and silicon appears
to be slightly enhanced compared to solar abundance.
The resolution of the RGS spectra is sufficient to measure the rotational
velocity of the bounday layer via Doppler broadening of emission lines.
We calculated the broadening using the O\,{\sc viii} emission line at 
19\AA\ which is the strongest line in the spectrum. We used a Gaussian
model to calculate the $\sigma$ of the line ($\sigma \times 2.4$=FWHM)
along with a power-law/bremmstrahlung for the continuum. We find that
FWHM=0.044\AA\ and the corresponding velocity in the line of sight is
695 km s$^{-1}$ ($\Delta \lambda$/$\lambda$ = $v/c$). We also checked the
resolution of RGS at around 19\AA\ and found 0.04-0.05 \AA\ and we 
caution that our measurement is on the limit of the RGS spectral 
resolution. 

\section{Discussion} 

The maximum X-ray shock temperature we obtain is 29-33 keV, based on the
EPIC MOS1, 2 and pn data, as they 
have higher count rates and braoder energy ranges than the RGS data.
RU Peg has a harder X-ray spectrum than in most dwarf novae, 
but softer than U Gem in quiescence, which also contains a massive
white dwarf \citep{sio98,lon99}. This result is consistent
with and confirms the large mass of the white dwarf in RU Peg.   

The X-ray luminosity of RU Peg is $4.1 \times 10^{32}$ ergs$~$s$^{-1}$ (0.2-10.0 keV),  
and assuming we see only half of the
boundary layer (if it is close to the star), we 
have $L_{bl} = 8.2 \times 10^{32}$ ergs$~$s$^{-1}$.  
This is in the range of the other quiescent DNe observed with XMM-Newton   
by \citet{pan05}. However, in order to fully compared RU Peg with these
other systems, we also need to consider the UV luminosity.  
For the UV we use the spectral luminosity at 290 nm $L_{UVW1}$
obtained from the OM data (see section 2.1).  
We have reproduced Figure 4 of \citet{pan05} in Figure 6 with the inclusion
of RU Peg, which shows the quiescent DNe observed with XMM-Newton plotted
on a $L_{UVW1}$ against $L_{bl}$ graph. 
In this figure the solid and dotted lines show the relation $L_{bl}=L_{disk}$ 
for a UV luminosity
predicted by a simple accretion disk model with an inner radius of 
5000km and 10,000km respectively. This simple disk model does not
include UV contribution from the WD, or outer edge of the BL where
it meets the inner disk. The location of a system in the vicinity of
this diagonal (e.g. SU UMa, WW Hyi) indicates that the boundary layer luminosity
$L_{bl}$ is comparable to the disk luminosity (here $L_{disk}=L_{UVW1}$).  
For RU Peg, as for U Gem for example, the excess of UV is due to the 
contribution from the WD marked on the left of the graph. 
RU Peg has a UV luminosity  
corresponding to 53,350 K for a 8,000 km radius WD (assuming the flux
scales simply as $\propto T^4$),   
or 75,450 K WD for a 4,000 km radius (more consistent with the large mass of 
RU Peg). This confirms the temperature of the WD as derived from 
the {\it FUSE} spectrum \citep{god08},  
and  puts RU Peg (literally) in line with all the
other quiescent DNe such as VW Hyi, U Gem, SU UMa, OY Car, and AB Dra,
on the $L_{UVW1}$ versus $L_{bl}$ graph.  

Because of its large mass (and therefore small radius),  
the WD in RU Peg has a deeper potential well, and v$_{\rm Kep}$ of the
order of 5-6,000 km$~$s$^{-1}$ in the BL.    
As the matter is decelerated in the BL, the X-rays emitting gas 
has velocities of a few 1000 km$~$s$^{-1}$.   
The modeling of the O\,{\sc viii} emission line at 19\AA\  implies a projected 
rotational broadening $v_{rot} \sin{i}=$ 695 km$~$s$^{-1}$, 
i.e. the line is emitted from
material rotating at $\sim$ 936-1245 km/$~$s$^{-1}$ (since 
$i \sim 34^{\circ}-48^{\circ}$) or about 1/6 of the Keplerian speed. 
This velocity is still much larger than the rotation speed of the white dwarf
inferred from the {\it FUSE} spectrum (40 km ~s$^{-1}$ ;
\citep{god08}). 
This implies that  
the X-ray emission comes directly from the decelerating boundary layer material.
It is possible that the X-ray emission originates 
in the equatorial region of the white dwarf as shown by \citet{pir04}, who 
found that poloidal motion may be negligible at low mass accretion rates 
as a characteristics of DNe in quiescence.
In such a case most of the dissipated energy is radiated
back into the disk, which may justify the use of a one-dimensional 
treatment (such as \citet{nar93,pop99}). 

The X-ray luminosity we computed corresponds to a boundary layer luminosity
(eq.2) 
for a mass accretion rate of $2.0 \times 10^{-11}M_{\odot}$/yr assuming a 
$1.29M_{\odot}$ WD mass, and it increases to 
$3.0 \times 10^{-11}M_{\odot}$/yr  for a $1.2M_{\odot}$ WD mass. 
This is entirely   
consistent with a quiescent accretion rate. 

We checked the correlation between the variability of the X-ray data 
and the UV data and found two components; one delayed and the other undelayed.
The X-ray and UV emission originate from distinct regions in the 
binary. 
Since RU Peg is a non-magnetic system, 
the X-ray emission is entirely from the inner edge of 
the boundary layer very close to the WD surface.
As to the UV radiation, it is emitted by the heated WD, the very inner 
disk, and also that region where the 
outer boundary layer meets the disk \citep{pop99}.  
The significant modulation correlation at $\Delta t$$\sim$0 lag is expected to 
be caused by reprocessing of X-rays (i.e., irradiation by X-rays) in the 
accretion disk. Such time lags are on the order of milliseconds and 
proportional to light travel time which is well beyond the time resolution
in our light curves. The delayed component of $\Delta t$$\sim$116 sec lag
is much longer and can not be produced by light travel effects nor
by reprocessing of the X-ray, since the X-ray trails behind the UV.
The only viable explanation, is that the time lag is the time it takes
for matter to move inwards from the very inner disk (emitting in the UV)
onto the stellar surface (emitting in the X-ray). This is the time 
it takes to spin down the material in the boundary layer  
$\tau_{spin}=\Delta t=116$s . 
The modulations of the UV component and its lagging X-ray counterpart are due
to modulations in $\dot{M}$.  
A comparable time delayed ($\Delta t$$\sim$100 sec) component
was also detected for VW Hyi \citep{pan03}, and a much shorter one
($\sim 7$s; Revnivtsev et al. 2011) was detected for the intermediate polar EX Hya (indicating
that the transit of matter through magnetic field lines to the poles
is faster than though the non-magnetic boundary layer).  

Following the work of \citet{god05}, we use the 
spin down time $\tau_{spin}=116$s and the 
rotation (or dynamical) time    
$\tau_{rot}=2 \pi r /v_K(r)$=25.3s to derive the 
viscous time in the BL $\tau_{\nu}= \tau^2_{spin}/\tau_{rot}=$532s. 
This gives a boundary layer viscosity 
$\nu = \delta_{bl}^2/\tau_{\nu}=4.8 \times 10^{13}$cm$^2$s$^{-1}$,
where we have assumed $M_{wd}=1.29M_{\odot}$, $R_{wd}=4,000$km, 
and a boundary layer size $\delta_{bl}$ given by the 
boundary layer radius 
$r_{bl}= (1+\delta_{bl})R_{wd}=1.4R_{wd}$ \citep{pop99}.   
In the alpha viscosity prescription $\nu = \alpha c_s H$, 
the value of the alpha parameter is then simply 
$\alpha = \nu / (c_s H) \approx 0.003$, where we assumed 
$c_s \approx 10^8$cm$~$s$^{-1}$ (for $T\approx 10^8$K) and 
the vertical thickness of the boundary layer $H \approx 0.4R_{wd}$
\citep{pop99}. This value of $\alpha$ for the BL of RU Peg 
and the one derived for the BL of VW Hyi (0.009 - \citet{god05}) 
indicate that the alpha viscosity parameter in the BL ($\alpha_{bl}$)
is much smaller than in the disk ($\alpha_{disk}$; 
at least in quiescent dwarf novae). This is consistent with the         
analytical estimates of \citet{sha88} and \citet{god95a} which
obtained $\alpha_{bl} << \alpha_{disk}$. 
Since the source of the boundary layer viscosity 
is unknown \citep{ino99,pop01}, this result is important for future 
theoretical work investigating the source of the viscosity
in the boundary layer.

\acknowledgments

PG wishes to thank 
Bill Blair for his kind hospitality at the Johns
Hopkins University where part of this work was carried out. 
Except for the symbol and temperature mark of RU Peg, Figure 
6 was taken from \citet{pan05}, who kindly agreed that 
we reproduce Figure 4 from their original work.  
This work is  based on observations obtained with the
XMM-Newton, an ESA science mission with instruments
and contributions directely funded by ESA member states
and by NASA.  
Support for this work was provided by NASA through
grant numbers NNX08AX43G (XMM-Newton AO7) 
to Villanova University.

\clearpage

\begin{table} 
\caption{XMM-Newton Observations of RU Peg} 
\vspace{0.5cm} 
\begin{tabular}{lc}
\hline
Instument & Count Rate  \\ 
          & $<s^{-1}>$    \\ 
\hline
RGS2      & $0.2881 \pm 0.0030$  \\ 
RGS1      & $0.2351 \pm 0.0029$  \\ 
EPIC pn   & $10.03  \pm 0.017 $  \\ 
EPIC MOS1 & $3.154  \pm 0.0084$  \\ 
EPIC MOS2 & $3.241  \pm 0.0085$  \\ 
\hline
\end{tabular}
\end{table}

\begin{table}
\begin{center}
\caption{Spectral parameters of fit to the combined EPIC spectrum of RU Peg in
the energy range 0.2-10 keV. $N_{H}$ is the absorbing column,
$\alpha$ is the index for power-law emissivity function
($dEM = (T/T_{max})^{\alpha-1} dT/T_{max}$),
$T_{max}$ is the maximum temperature for CEVMKL model.
Element names stand for the abundance relative to solar abundances, 
Gaussian LineE is the line center for the emission
line, $\sigma_{G}$ is the line width; 
$K_{CEVMKL}$ and $K_{G}$ are the normalizations 
for CEVMKL and Gaussian models respectively. 
The unabsorbed X-ray flux is given
in the range 0.2-10.0 keV.
All error ranges are given in  90\% confidence level
($\Delta \chi^2$=2.71 for a single parameter) 
}
\vspace{0.5cm} 
\begin{tabular}{@{}lrrrrlrlr@{}}
\hline
\hline
\multicolumn{1}{l}{PARAMETER} &
\multicolumn{1}{r}{VALUE} \\
\hline
$N_{H}$ ($10^{22}$ atoms/cm$^{2}$) & 0.044$^{+0.001}_{-0.001}$ \\
$\alpha$ & 1.05$^{+0.03}_{-0.03}$  \\
$T_{max}$ (keV) & 31.7$^{+1.7}_{-2.0}$  \\
O  & 0.3$^{+0.06}_{-0.06}$  \\
Ne & 0.55$^{+0.16}_{-0.16}$  \\
Mg & 1.3$^{+0.2}_{-0.2}$  \\
Si & 0.8$^{+0.14}_{-0.14}$  \\
S & 0.9$^{+0.3}_{-0.2}$  \\
Ca & 1.8$^{+0.8}_{-0.7}$  \\
Fe & 0.8$^{+0.04}_{-0.04}$  \\
$K_{CEVMKL}$ & 0.047$^{+0.001}_{-0.001}$ \\
Gaussian LineE (keV) & 6.4 (fixed) \\
$\sigma_{G}$  (keV) & 0.19$^{+0.025}_{-0.025}$   \\
$K_{G}$ & 0.000052$^{+0.000006}_{-0.000006}$  \\
Flux ($10^{-11}$ ergs$~$cm$^{-2}$s$^{-1}$) & 4.1$^{+0.2}_{-0.2}$  \\
$\chi^2_{\nu}$ (d.o.f.) & 1.5 (1741) \\
\hline
\end{tabular}
\end{center}
\end{table}

\begin{table} 
\caption{Line Identifications of the XMM-Newton RGS spectrum of RU Peg} 
\vspace{0.5cm} 
\begin{tabular}{lr}
\hline
Ion              & Wavelength  \\ 
                 &  $< \AA >$    \\ 
\hline
Mg\,{\sc xii}    &  7.20      \\ 
                 &  7.45      \\ 
Ne\,{\sc x}      & 12.15      \\ 
Fe\,{\sc xvii}   & 15.03      \\ 
O\,{\sc viii}    & 16.05      \\ 
Fe\,{\sc xvii}   & 16.80      \\ 
Fe\,{\sc xvii}   & 17.07      \\ 
O\,{\sc viii}    & 19.00      \\ 
O\,{\sc vii}(i)  & 21.80      \\ 
O\,{\sc vii}(f)  & 22.11      \\ 
N\,{\sc vii}     & 24.77      \\ 
\hline
\end{tabular}
\end{table}

\begin{table}
\begin{center}
\caption{Spectral parameters of the fit to the RGS spectrum of  RU Peg in
the energy range 0.2-2.5 keV with same definitions as in Table 2. 
}
\vspace{0.5cm} 
\begin{tabular}{@{}lrrrrlrlr@{}}
\hline
\hline
\multicolumn{1}{l}{PARAMETER} &
\multicolumn{1}{r}{VALUE} \\
\hline
$N_{H}$ ($10^{22}$ atoms/cm$^{2}$) & 0.031$^{+0.007}_{-0.007}$ \\
$\alpha$ & 1.2$^{+0.16}_{-0.16}$  \\
$T_{max}$ (keV) & 24.1$^{+17.0}_{-7.0}$  \\
O  & 0.8$^{+0.13}_{-0.12}$  \\
Ne & 1.1$^{+0.5}_{-0.5}$  \\
Mg & 2.7$^{+1.2}_{-1.1}$  \\
Si & 3.4$^{+2.0}_{-2.0}$  \\
S &  1.4$^{<}_{-3.2}$  \\
Ca & 1.0 (fixed) \\
Fe & 1.0 (fixed)  \\
$K_{CEVMKL}$ & 0.043$^{+0.004}_{-0.005}$ \\
Flux ($10^{-11}$ ergs$~$cm$^{-2}$) & 3.9$^{+0.4}_{-0.4}$  \\
$\chi^2_{\nu}$ (d.o.f.) & 1.3 (283) \\
\hline
\end{tabular}
\end{center}
\end{table}

\begin{figure}
\vspace{-5.cm} 
\plotone{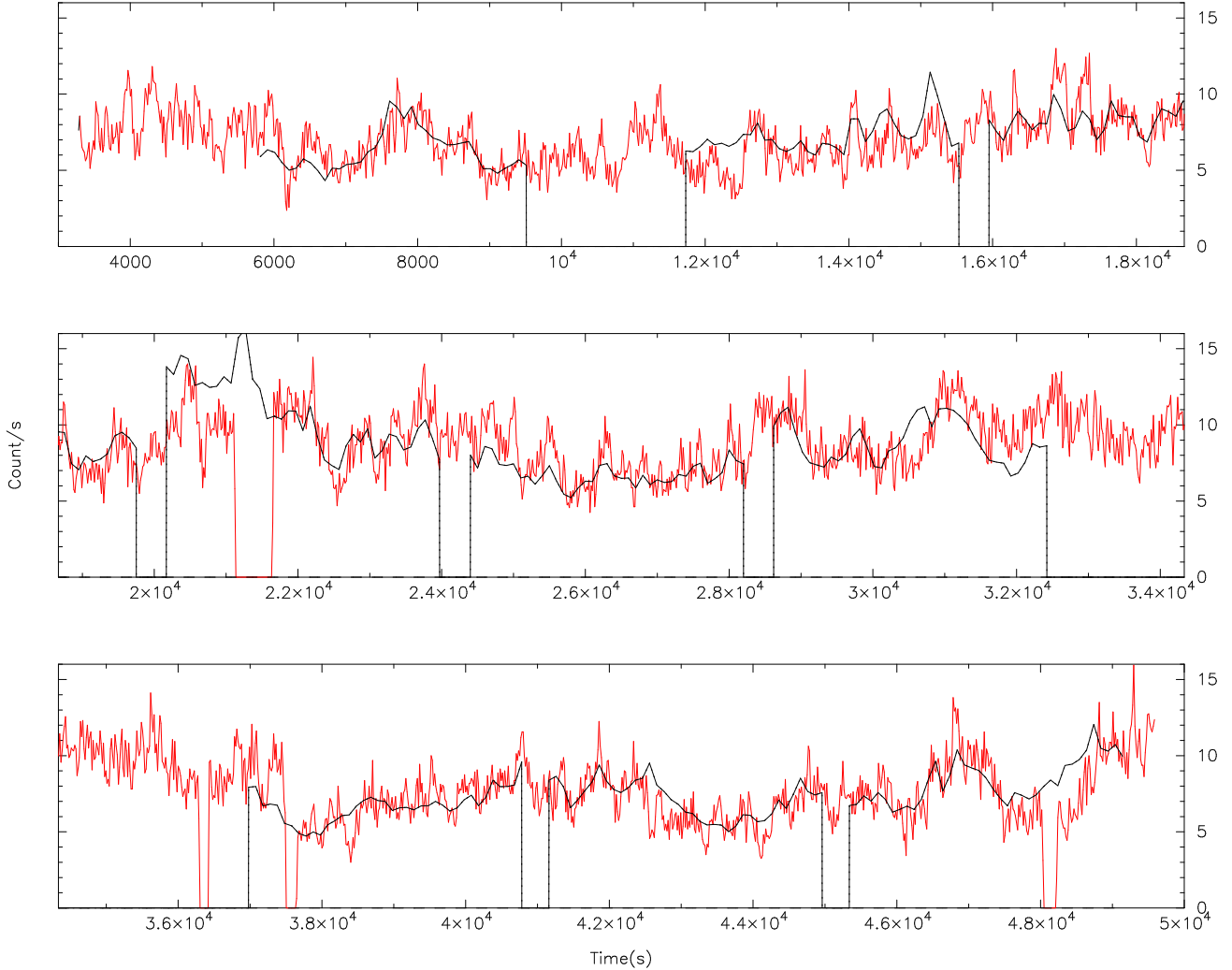}             
\caption{The EPIC X-ray light curve (binned at 16 s, in red) 
together with the OM UV light curve 
(binned here at 100 s for display, in black; the data were binned at 20 s for the
correlation check).
The count rate of the UV data has been divided by 
9.1736 to fit the count rate level of the 
X-ray data for easier comparison. The time $t=0$ 
corresponds to $t=3.29389320 \times 10^8$s, counted from the
MJD reference day 50,814, namely: 54,626.376 MJD.  
The time modulation of the UV data follows closely the 
time modulation of the X-ray data except around 
$t\approx $12 ks, 20-22 ks, 31 ks, \& 48.5 ks, where the
UV has {\it relatively} more flux.} 
\end{figure}

\begin{figure}
\vspace{-5.cm}
\plottwo{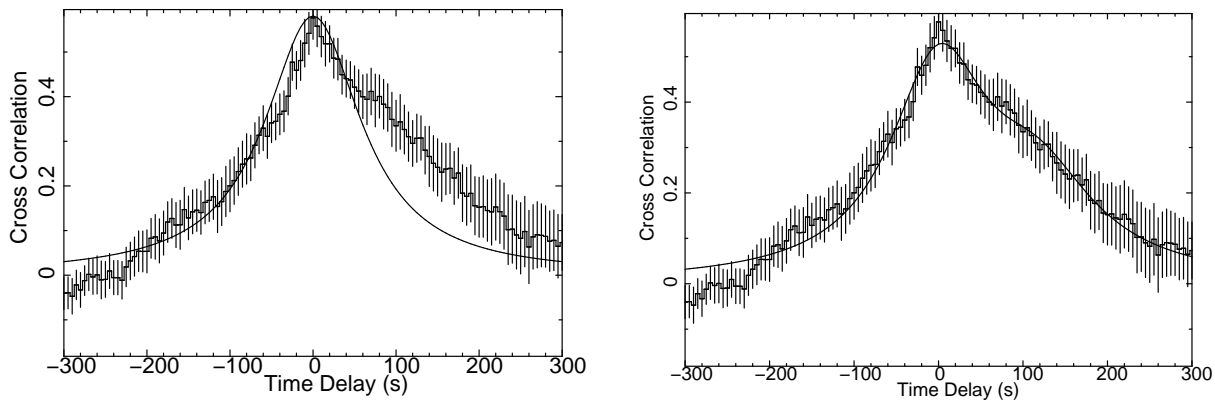}{f2b.ps}
\caption{Cross-correlation between X-ray and UV light curves. 
Left: a single Lorentzian is used to fit the data at 0; 
Right: a second Lorentzian fit is added to fit the delayed component. 
The figures 
show strong correlation at zero time lag.
The asymmetric profile centered at zero time lag (asymmetry towards positive 
lag) indicates that the X-ray variations are delayed with respect to the UV 
variations. Error bars indicate the standard deviations from the average of the
value of the cross-correlation at any given lag in different time segments where
power spectra are calculated.
}

\end{figure}

\begin{figure}
\vspace{-5.cm} 
\plotone{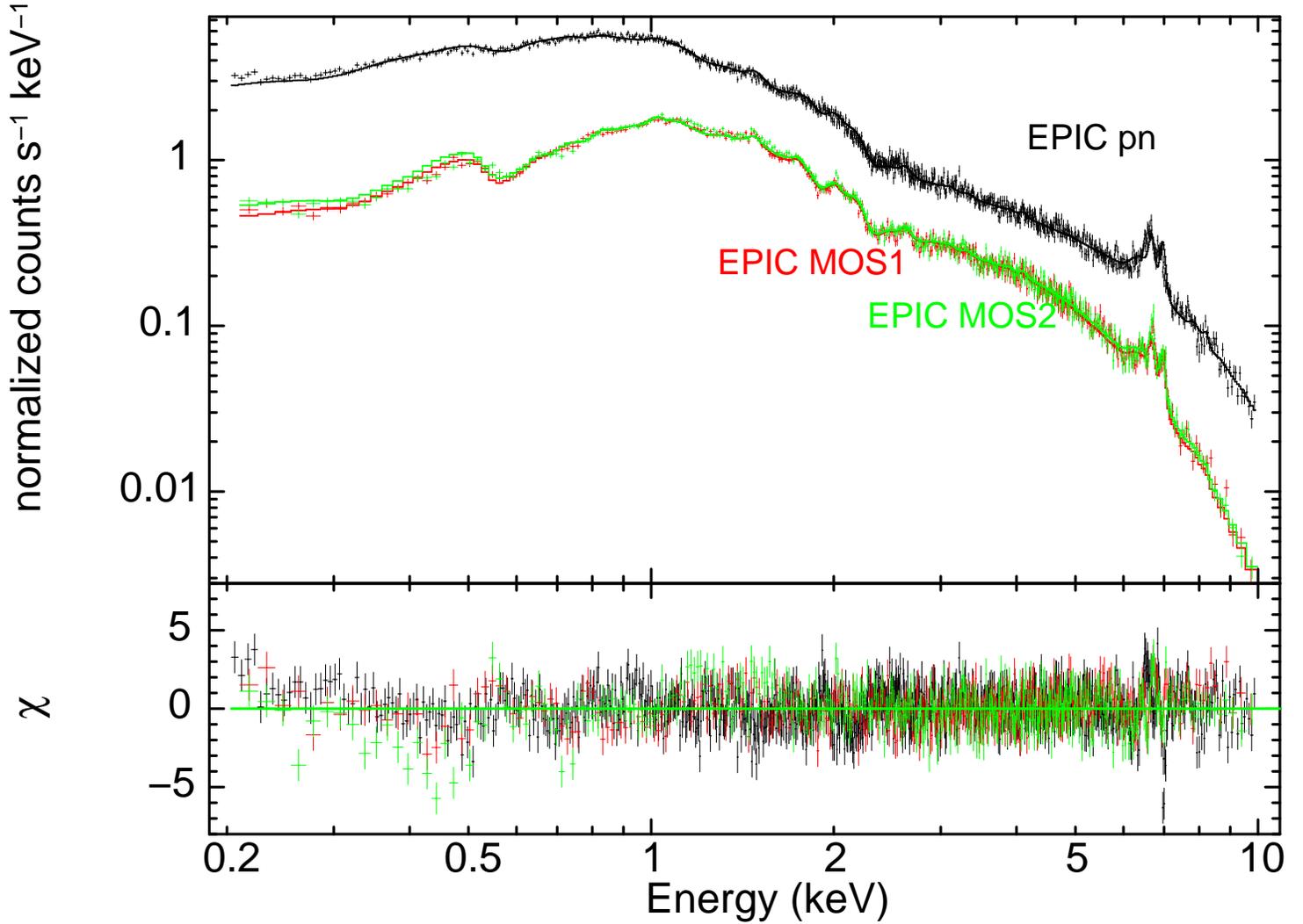}                  
\caption{
EPIC MOS1,2 and pn spectra simultaneously fitted to the same model 
TBabs$\times$CEVMKL. 
The model includes an ISM absorption model
and a multi-temperature plasma emission model built from the MEKAL 
code.
The maximum shock temperature obtained from the fit is 
31.7 keV. The second panel below the fitted spectra is the residuals
in standard deviations. One can notice a slight CTI problem of the pn
detector around the iron lines (6.4-6.9 keV) where the fluctuations in the
residuals due to CTI are evident. This does not affect the global 
fitting procedure but only slightly increases the reduced $\chi^2$ of the fits. 
}  
\end{figure}

\begin{figure}
\vspace{-5.cm} 
\plotone{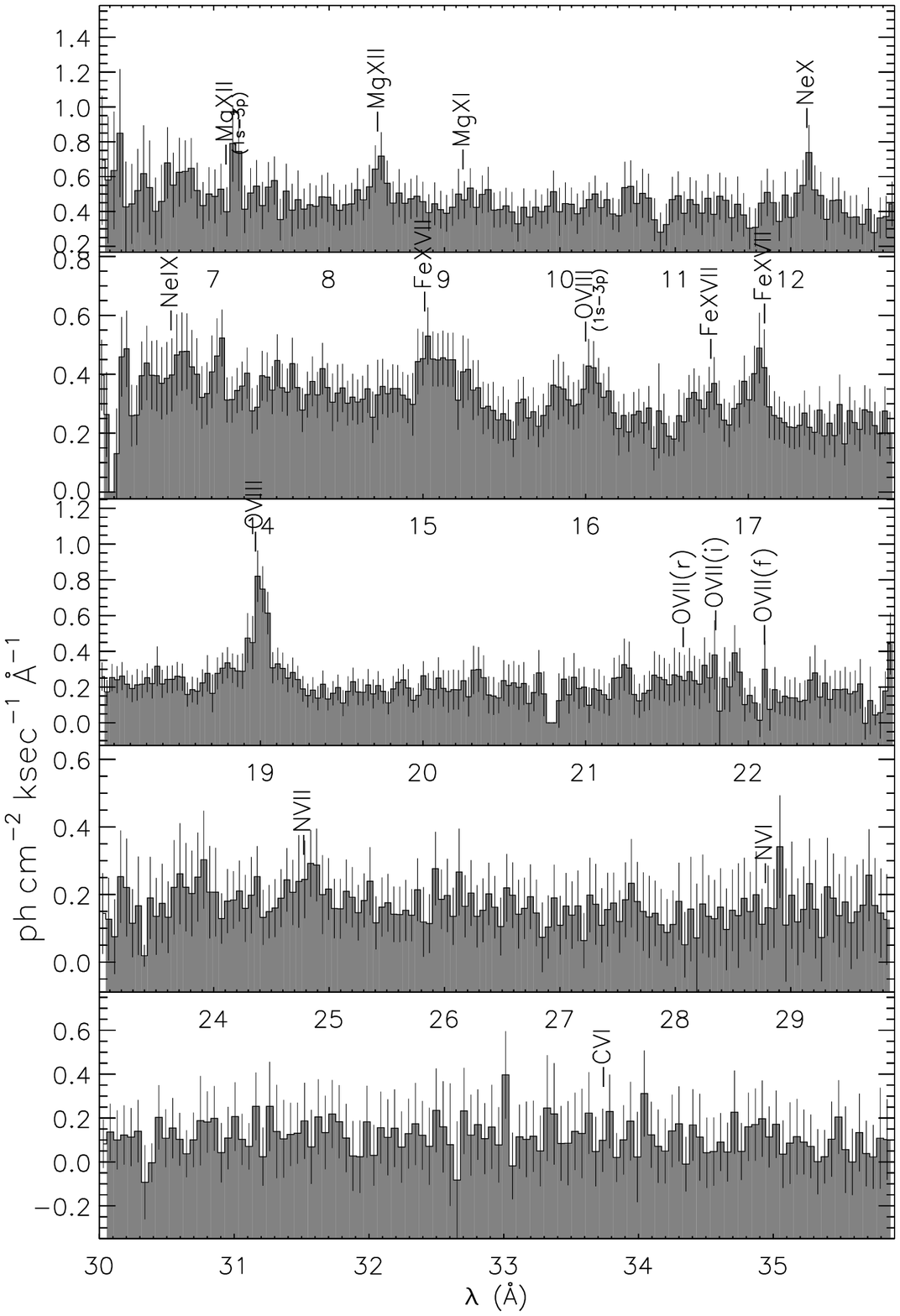}                  
\caption{The fluxed RGS Spectrum of RU Peg with line identifications.} 
\end{figure}

\begin{figure}
\vspace{-5.cm} 
\plotone{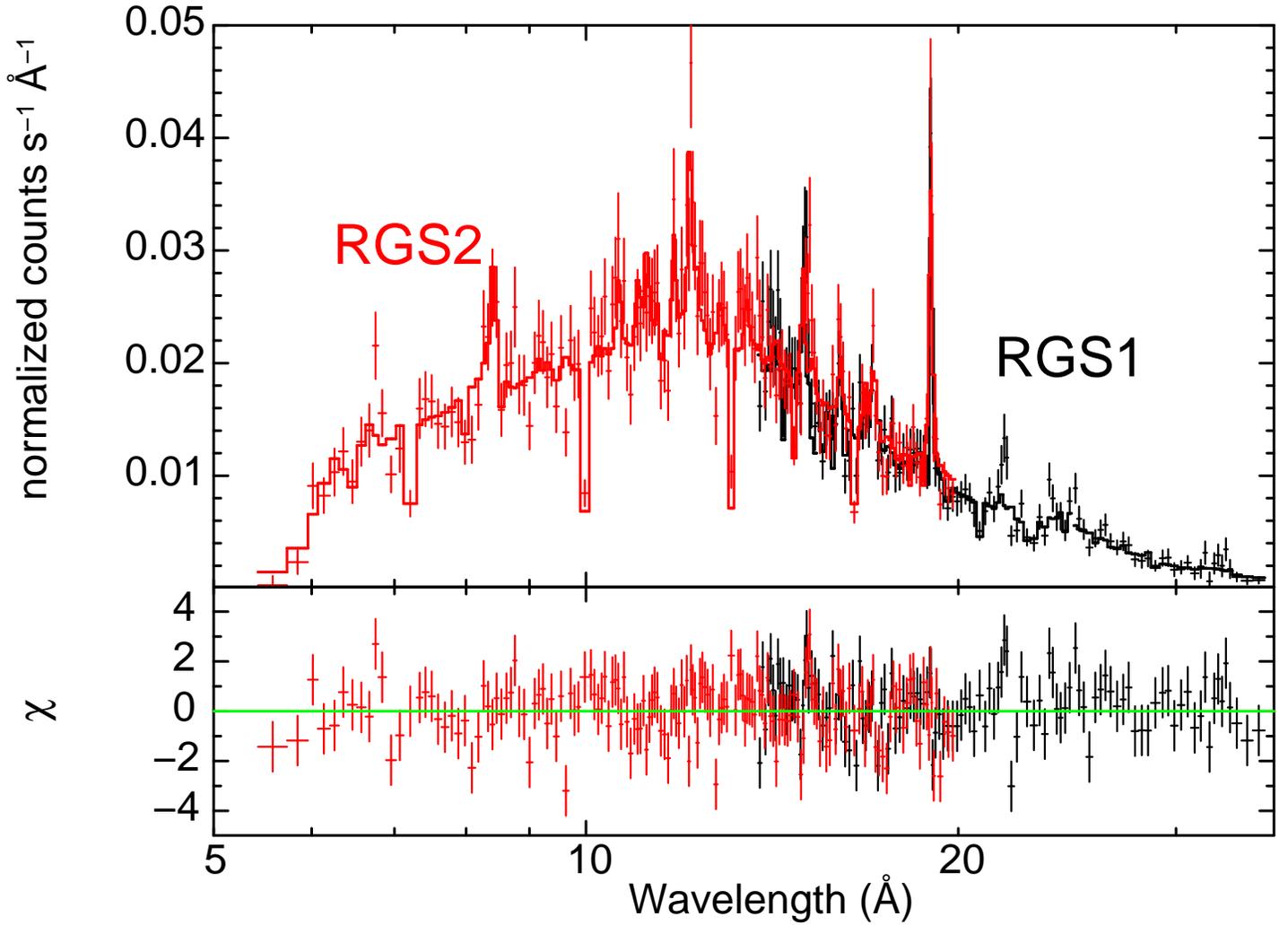}                  
\caption{
The simultaneously fitted RGS 1 and RGS 2 spectrum of RU Peg.       
The same model TBabs$\times$CEVMKL is used for the fit. 
The residuals in the second panel below are
in standard deviations. 
} 
\end{figure}

\begin{figure}
\plotone{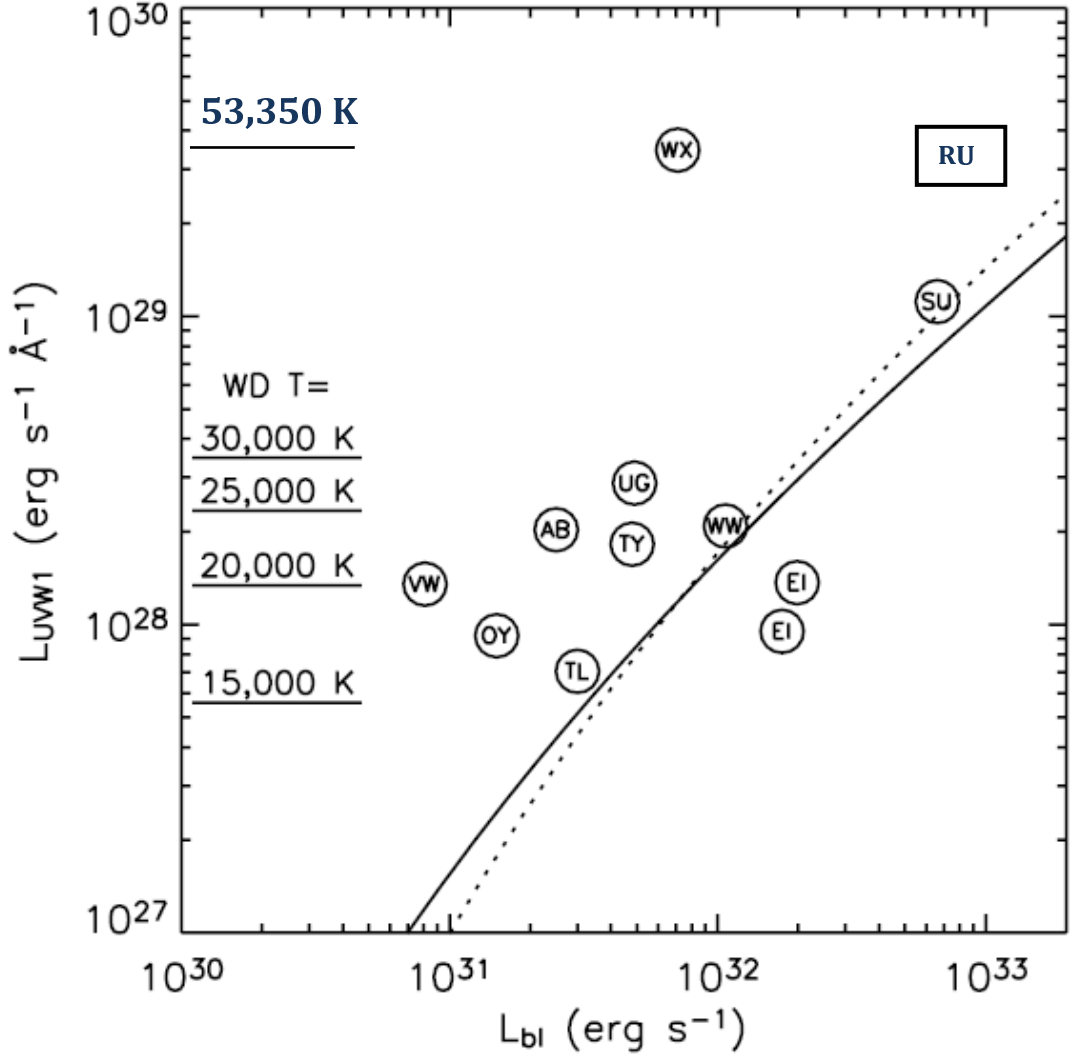} 
\caption{
The boundary layer (X-ray) luminosities
vs. the UV spectral luminosity at 290 nm, as derived by    
\citet{pan05} for quiescent DNe observed with XMM-Newton.
We have added RU Peg on the graph.  
The solid and dotted lines show $L_{bl}=L_{disk}(=L_{UVW1})$ 
assuming UV luminosity predicted by a simple
accretion disk model for an inner radius $R_{in}=5000$ km and 
10,000 km respectively. For systems higher above the line, such 
as RU Peg, the UV excess is due to the contribution from the WD.  
The UV luminosity of a WD with a 8000 km
radius at various temperature is shown on the left. RU Peg corresponds
here to a $\sim$53,350 K WD with a radius of 8000 km (which we have also
added to the original graph), which translates
to a $\sim$75,000 K for a 4000 km WD, more in line with the large mass
of the WD in RU Peg. WX Hyi was caught in outburst explaining its UV excess.  
}  
\end{figure}

\end{document}